# The $T{=}0$ neutron-proton pairing correlations in the superdeformed rotational bands around $^{60}$Zn

J. Dobaczewski,[1,2] J. Dudek,[2] and R. Wyss[3]

[1]*Institute of Theoretical Physics, Warsaw University, Hoża 69, PL-00681, Warsaw, Poland*
[2]*Institut de Recherches Subatomiques, UMR7500, CNRS-IN2P3 and Université Louis Pasteur,
F-67037 Strasbourg Cedex 2, France*
[3]*KTH-Kärnfysik, Frescativ. 24, S-104 05 Stockholm, Sweden*

The superdeformed bands in $^{58}$Cu, $^{59}$Cu, $^{60}$Zn, and $^{61}$Zn are analyzed within the frameworks of the Skyrme-Hartree-Fock as well as Strutinsky-Woods-Saxon total routhian surface methods with and without the $T{=}1$ pairing correlations. It is shown that a consistent description within these standard approaches cannot be achieved. A $T{=}0$ neutron-proton pairing configuration mixing of signature-separated bands in $^{60}$Zn is suggested as a possible solution to the problem.

## I. INTRODUCTION

A possibility to find signatures of the $T{=}0$ neutron-proton (n-p) pairing correlations in $N{\sim}Z$ nuclei is recently a subject of a significant fraction of experimental and theoretical studies in nuclear structure physics. At low spins, such correlations allow for a consistent description of ground states and low-$T$ excitations in even-even and odd-odd $N{=}Z$ nuclei [1]. This type of correlations may also be, in principle, visible through changes in structure of rotational nuclear bands. For example, the significance of the so-called delayed alignments in $N{=}Z$ nuclei is at present intensely investigated, both experimentally [2,3] and theoretically, e.g., see recent Refs. [4–8] and references cited therein.

In the present study we address another experimental fact which may constitute such a signature, namely, an anomalous behavior of the second moment of inertia $\mathcal{J}^{(2)}$ in the superdeformed (SD) band of $^{60}$Zn, as compared to its neighbors. The peak of $\mathcal{J}^{(2)}$ observed at low spins in $^{60}$Zn has been in the original experimental paper [9] tentatively interpreted as the simultaneous alignment of the $T{=}1$ pairs of $g_{9/2}$ protons and neutrons, although no calculation supporting such a hypothesis was presented. Together with the discovery of the analogous SD band in $^{61}$Zn [10], where only a small bump of $\mathcal{J}^{(2)}$ was observed, the $T{=}0$ paired band crossing was proposed as an underlying structure of the $^{60}$Zn band. Indeed, in a simple scenario such a crossing would be entirely blocked in $^{61}$Zn, while for the $T{=}1$ pairing only the neutron crossing would be blocked, leaving half of the peak intact. The $T{=}0$ paired-band structure was further corroborated by the lack of the analogous peak in the SD band in $^{59}$Cu [11].

On the other hand, the $T{=}1$ pairing calculations performed in Ref. [12] indeed resulted in a strong rise of $\mathcal{J}^{(2)}$ with decreasing angular frequency of the $^{60}$Zn SD band. However, at lower frequencies solutions could not have been obtained, and hence the complete peak of $\mathcal{J}^{(2)}$ was not reproduced. Neither the blocked calculations in neighboring odd and odd-odd nuclei were performed to support the possibility of reproducing smooth SD bands there within the $T{=}1$ pairing scenario. It was only argued that deformation effects can be important for the complete understanding of the physical picture.

In this study we present the first set of consistent calculations of the SD bands in $^{58}$Cu, $^{59}$Cu, $^{60}$Zn, and $^{61}$Zn, performed within the $T{=}1$ pairing hypothesis. We show that the simple scenario of blocking either the neutron or proton $T{=}1$ pair indeed does not hold, and a more complicated picture is obtained. However, a gradual disappearance of the $T{=}1$ pairing correlations with increasing rotational frequency always creates too large values of $\mathcal{J}^{(2)}$ at high frequencies, in disagreement with data. In fact at high frequencies the values of $\mathcal{J}^{(2)}$, as well as the values of relative alignments, are perfectly well described by calculations that altogether neglect the $T{=}1$ pairing correlations. Therefore, it seems that the only effect that the no-pairing theory cannot describe is the peak of $\mathcal{J}^{(2)}$ in $^{60}$Zn. Therefore, we attempt to describe this structure by a simple $T{=}0$ n-p pairing configuration mixing of unpaired solutions.

The superdeformed (SD) bands in the $A{\simeq}60$ nuclei have already been studied theoretically within various approaches [13–21]. In the present paper we use two methods: (i) the cranked Hartree-Fock (HF) method, solved by using the HFODD (v1.75r) computer code [22], with the Skyrme SLy4 [23] effective interaction and no pairing (see Ref. [21] for details), and (ii) the cranked Strutinsky total routhian surface (TRS) calculations based on a deformed Woods-Saxon (WS) potential [24], with the $T{=}1$ pairing correlations treated within the approximate particle number projection by means of the Lipkin-Nogami (LN) method (see Refs. [25–27] for details). Results of these calculations are presented in Secs. II and III, respectively, while in Sec. IV we present the $T{=}0$ n-p pairing configuration-mixing calculations based on the HF results.

## II. HARTREE-FOCK CALCULATIONS (NO PAIRING)

Before discussing the SD bands in nuclei around $^{60}$Zn, we briefly present some generic features of the corresponding single-particle spectra. The HF neutron single-particle orbitals near the SD $N=Z=30$ magic gap, calculated in $^{60}$Zn, are shown in Fig. 1. For protons the corresponding Routhian diagrams are almost identical apart from a uniform shift in energy. The single-particle spectra show large gaps at $N(Z)=30$ that are stable up to the highest frequencies. At the bottom of the SD magic gap there appear two strongly deformation-driving intruder orbitals $[440]1/2(r=\pm i)$, that originate from the $N_0=4$ harmonic oscillator (HO) shell, or more specifically, from the spherical $1g_{9/2}$ subshell, and therefore are denoted as $4^1\equiv[440]1/2(r=-i)$ and $4^2\equiv[440]1/2(r=+i)$. Above the gap, one can see six low-lying orbitals, i.e., the next two intruder states $4^3\equiv[431]3/2(r=-i)$ and $4^4\equiv[431]3/2(r=+i)$, as well as four negative-parity orbitals which in the present study are denoted as $f_\pm\equiv[303]7/2(r=\pm i)$ and $p_\pm\equiv[310]1/2(r=\pm i)$. The $f_\pm$ orbitals are in fact the hole states originating from the $1f_{7/2}$ spherical subshell, while the $p_\pm$ orbitals are strong mixtures of the $1f$ and $2p$ spherical subshells, i.e., symbol $p_\pm$ is assigned only to fix a convenient naming convention.

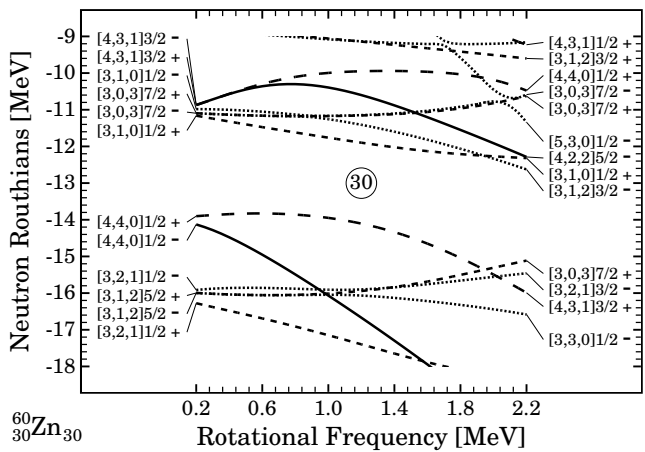

FIG. 1. Hartree-Fock neutron single-particle Routhians in the SD doubly magic configuration $4^24^2$ of $^{60}$Zn, calculated for the Skyrme interactions SLy4. Lines denoting the four (parity, signature) combination are: long-dashed $(+,+i)$, solid $(+,-i)$, short-dashed $(-,+i)$, and dotted $(-,-i)$. Standard Nilsson labels are determined by finding the dominating Nilsson components of the HF wave-functions at low (left set) and high (right set) rotational frequencies.

The doubly-magic SD configuration in $^{60}$Zn [9], denoted by $4^24^2$, corresponds to occupying all orbitals below the $N=Z=30$ gaps, and leaving empty all those that are above these gaps. Similarly, following the assignments of configurations proposed for experimentally observed bands, we have calculated three other SD bands, for the $4^14^1$ ($^{58}$Cu [28]), $4^24^1$ ($^{59}$Cu [11]), and $4^34^2$ ($^{61}$Zn [10]) configurations. The relative alignments (i.e., differences of angular momenta at fixed rotational frequencies) with respect to the SD band in $^{58}$Cu are shown in Fig. 2. Since the experimental SD bands in $^{59}$Cu, $^{60}$Zn, and $^{61}$Zn extend to higher rotational frequencies than that in $^{58}$Cu, we have artificially extended the latter band by adding two gamma rays at 3641 and 4128 keV. This was done for the presentation purpose only; alternatively, we could have used the $^{59}$Cu band as the reference, however, this would have not allowed us to show the relative alignments at lower rotational frequencies. Since the exit spins of the $^{58}$Cu and $^{61}$Zn bands have been measured only tentatively, in preparing Fig. 2 we have assumed the values of $I=9$ and $I=25/2\,\hbar$, respectively. In calculations, the angular momenta $I$ are identified with the average projections $\langle I_y \rangle$.

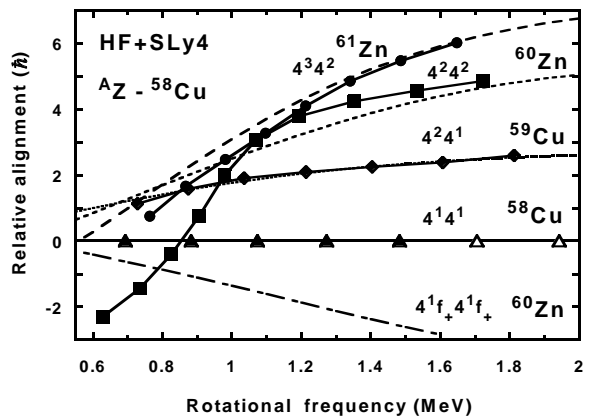

FIG. 2. Experimental [9,28,11,10] and calculated alignments of the SD bands in $^{59}$Cu (diamonds), $^{60}$Zn (squares), and $^{61}$Zn (circles), relative to the SD band in $^{58}$Cu (triangles). Calculations have been performed within the HF method with the SLy4 Skyrme interaction. In order to provide for a suitable reference at high rotational frequencies, two artificial gamma rays (open triangles) have been added to the $^{58}$Cu experimental data (see text).

In Fig. 3 we present a similar comparison between the measured and calculated dynamic moments of inertia $\mathcal{J}^{(2)}=\partial I(\omega)/\partial\omega$. In $^{58}$Cu, $^{59}$Cu, and $^{61}$Zn we obtain very good theoretical description of measured relative alignments and second moments. This gives us strong arguments in favor of the assigned configurations. However, unexpectedly, the SD band in the doubly-magic SD nucleus $^{60}$Zn deviates strongly from the theoretical predictions. This has been tentatively interpreted as an effect of the simultaneous alignment of the $g_{9/2}$ neutrons and protons [9], or as a manifestation of the $T=0$ n-p correlations [10]. In the present paper we perform the first calculations based on these two assumptions.

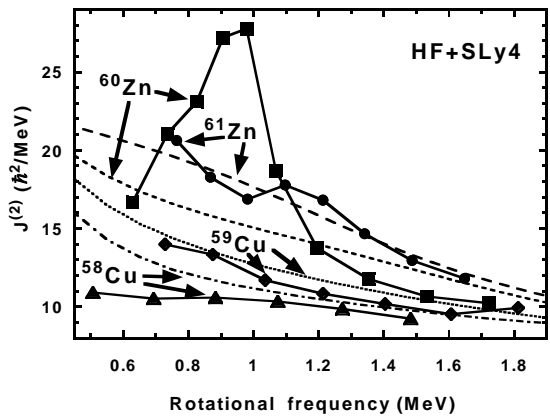

FIG. 3. Same as in Fig. 2 but for the second moments of inertia $\mathcal{J}^{(2)}$.

## III. STRUTINSKY CALCULATIONS WITH AND WITHOUT THE $T=1$ PAIRING CORRELATIONS

To shed more light especially on the role played by the $T=1$ pairing, we performed cranked Strutinsky type calculations based on a deformed WS potential. By comparing two sets of calculations, with and without the $T=1$ pairing, we aim at tracing the contribution and influence of the $T=1$ interaction. The $T=1$ pairing interaction is based on a seniority type force and a double stretched quadrupole interaction [29]. For the case of odd nucleon number and/or excited configurations, each configuration is blocked self-consistently [30]. The model has been successful in the description of rotational states in a wide range of nuclei.

To probe the sensitivity of our results to the macroscopic input, we performed two sets of calculations based on: (i) the Myers-Swiatecki liquid-drop (MS LD) mass formula [31] and (ii) the folded Yukawa plus exponential (FY) mass formula [32]. The MS LD mass formula can be considered as rather stiff towards deforming the nucleus. On the other hand, the FY mass formula, explicitly involving the finite range of the nuclear force and the diffuseness of the nuclear surface, results in a softer surface energy and gives larger $\beta_4$ deformations. For very light nuclei, the contribution to the surface energy can become unphysically large, but for the case of mass $A=60$ region, one is still on safe grounds.

In contrast to the MS LD results, for the FY mass formula all the four nuclei discussed here have $T=1$ paired stable minima at deformations that are comparable to those obtained without pairing, but at larger values of the hexadecapole deformation parameters. The difference in deformations between these nuclei result in distinctly different response to the rotating field. Starting with $^{58}$Cu, we do not observe any distinct difference between the MS LD and FY calculations. Also for the case of $^{60}$Zn, no big differences are obtained, although the crossing is somewhat sharper here in the FY case. The largest difference occurs for the case of $^{61}$Zn. Since the FY calculations yield the deformation that is larger than for $^{60}$Zn, the neutron $g_{9/2}$ alignment is becoming more smooth, resulting in a rather modest hump in $\mathcal{J}^{(2)}$. In what follows we concentrate on the results of the FY calculations.

In Fig. 4 we present the WS neutron single-particle orbitals near the SD $N=Z=30$ magic gap in $^{60}$Zn. Even though the HF and WS spectra presented in Figs. 1 and 4 have been calculated within so much different approaches, they present striking similarities. The equilibrium deformations of the SD shapes, calculated within the HF and Strutinsky approaches for the $^{58}$Cu, $^{59}$Cu, $^{60}$Zn, and $^{61}$Zn nuclei, are presented in Table I. The values obtained at $\hbar\omega=0$ and 1 MeV illustrate the degree of the rotational polarization occurring along the SD bands. Similarly, by comparing the values for the four nuclei one can see the effects of the multipole polarizations induced by the $g_{9/2}$ protons and neutrons, cf. Refs. [16,21].

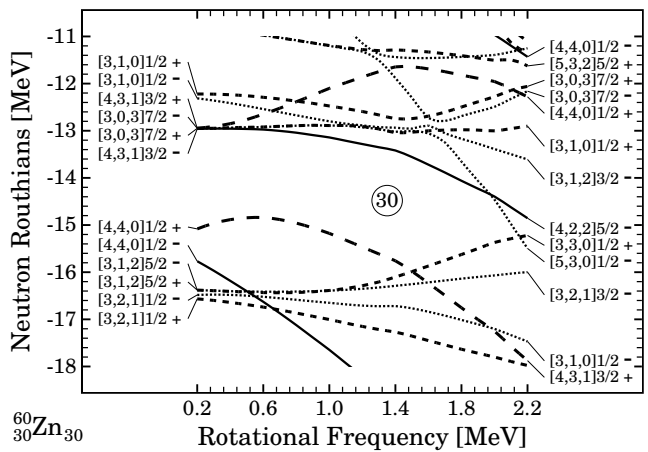

FIG. 4. Same as in Fig. 1 but for the Woods-Saxon potential with TRS deformations $\beta_2$, $\gamma$, and $\beta_4$ calculated along the $T=1$ paired SD band in $^{60}$Zn.

The LN method has been shown to be reliable for calculations of high spin states [25,26]. However, in the regime of a very weak pairing, one may encounter numerical problems in finding a proper solution. Indeed, this is the case for the present investigation, where starting from $\hbar\omega \approx 1.5$ MeV, the static pairing field essentially vanishes, and the pairing gaps become of the order of 100–200 keV. We employ two possible schemes to avoid a numerical break down of the paired solution. Either we fix the lowest value of the gap parameter to 100–200 keV, when no solution is found, or we make a transition to the non-pairing calculations. Since the calculations are done on a grid in deformation space, the frequency where the pairing solution encounters problems differs from point to point, giving fluctuations in the total energy. Changes in energy of the order of 50 keV are sufficient to cause oscillations in the calculated moments of inertia. In order to address the underlying physics, therefore, we smoothed the moments of inertia in the frequency range where such oscillations occur.

The resulting relative alignments and moments of inertia are depicted in Figs. 5 and 6, respectively. As can bee seen, the strong bump in the experimental moments of inertia of $^{60}$Zn is rather well reproduced in the WS+LN calculations with $T=1$ pairing. It indeed results from the alignment of a pair of $g_{9/2}$ protons and neutrons. The crossing frequency is somewhat too small in the calculations, but this can be considered as a detail in this context. However, it might also reflect the situation in heavier nuclei, where a similar shift has been observed [5,6] and attributed to the lack of the $T=0$ pairing. Note that a similar behavior of $\mathcal{J}^{(2)}$ has been also obtained in the relativistic-mean-field LN calculations of Ref. [12], although the increase of $\mathcal{J}^{(2)}$ at low frequencies could not have been obtained there.

TABLE I. Quadrupole ($\beta_2$ and $\gamma$) and hexadecapole ($\beta_4$) deformation parameters calculated for the SD configurations at $\hbar\omega=0$ and $1$ MeV. For each nucleus the three lines give: (a) the HF values obtained from the mass multipole moments $Q_{20}$, $Q_{22}$, and $Q_{40}$ through the second-order expressions for equivalent shapes [33,34], (b) the WS potential equilibrium deformations obtained by neglecting the pairing correlations, and (c) the WS potential equilibrium deformations obtained for the LN $T=1$ pairing. In the latter case, the $\hbar\omega=0$ MeV solution in $^{59}$Cu cannot be obtained.

| Nucleus | | $\hbar\omega=0$ MeV | | | $\hbar\omega=1$ MeV | | |
|---|---|---|---|---|---|---|---|
| | | $\beta_2$ | $\gamma$ | $\beta_4$ | $\beta_2$ | $\gamma$ | $\beta_4$ |
| $^{58}$Cu | (a) | 0.371 | $-1°$ | 0.051 | 0.343 | $5°$ | 0.037 |
| | (b) | 0.392 | $0°$ | 0.038 | 0.347 | $7°$ | 0.029 |
| | (c) | 0.374 | $-1°$ | 0.061 | 0.357 | $6°$ | 0.024 |
| $^{59}$Cu | (a) | 0.394 | $0°$ | 0.096 | 0.368 | $3°$ | 0.055 |
| | (b) | 0.429 | $0°$ | 0.066 | 0.377 | $5°$ | 0.038 |
| | (c) | — | — | — | 0.402 | $3°$ | 0.058 |
| $^{60}$Zn | (a) | 0.412 | $0°$ | 0.144 | 0.391 | $2°$ | 0.089 |
| | (b) | 0.453 | $0°$ | 0.088 | 0.418 | $3°$ | 0.058 |
| | (c) | 0.458 | $4°$ | 0.154 | 0.426 | $2°$ | 0.089 |
| $^{61}$Zn | (a) | 0.428 | $0°$ | 0.143 | 0.410 | $2°$ | 0.098 |
| | (b) | 0.468 | $4°$ | 0.092 | 0.445 | $2°$ | 0.067 |
| | (c) | 0.463 | $-1°$ | 0.123 | 0.418 | $-1°$ | 0.085 |

The experimental moment of inertia of $^{58}$Cu is totally flat, as one would expect since this crossing is blocked. At higher frequencies, the calculated $\mathcal{J}^{(2)}$ moment rises, resulting in a smaller hump centered at $\hbar\omega \approx 1.5$ MeV which is absent in the data. In self-consistent calculations, it is often difficult to exactly point to the cause of such apparent alignment as in the case of $^{58}$Cu. The dominant contribution appears to come from the rather sudden drop in pairing energy, where in the region of $\hbar\omega=1.4$–$1.6$ MeV, the pairing gap drops from a value of 0.4 MeV to essentially zero. At lower frequencies, the change in pairing correlations due to the Coriolis anti-pairing is of the order of 50 keV per step in $\hbar\omega(=0.1)$ MeV. The sudden drop in pairing energy results in a change in the routhian $dE_\omega$, giving rise to this apparent alignment. Thus for all nuclei calculated in this study with the $T=1$ pairing, there is an excess in the moments of inertia $\mathcal{J}^{(2)}$ at high frequencies, most pronounced in $^{58}$Cu. Such an excess is clearly absent in the experimental data. Since $^{58}$Cu is taken as a reference for the relative alignment in Fig. 5b, the excess of alignment obtained in this nucleus for the $T=1$ paired calculations perturbs relative alignments shown for other nuclei. Another reference nucleus, e.g., $^{60}$Zn may yield a better agreement with experiment.

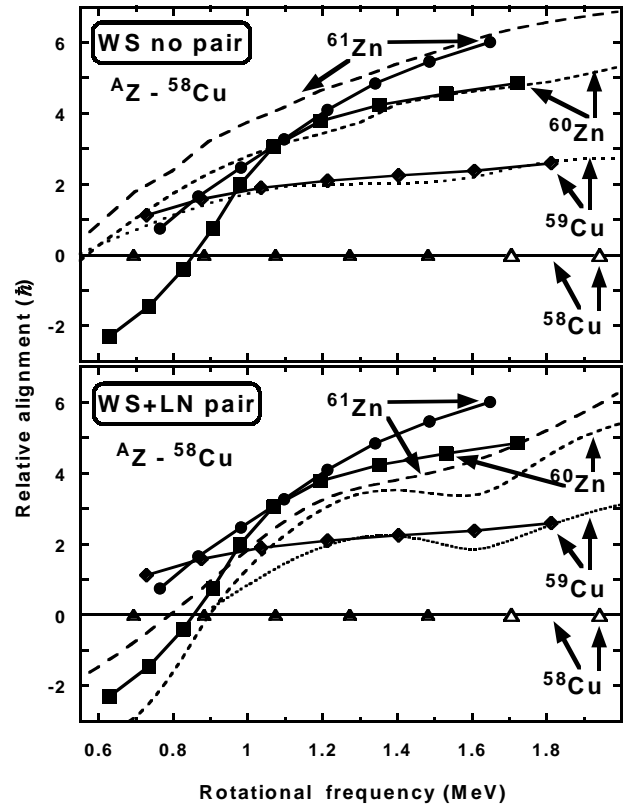

FIG. 5. Same as in Fig. 2 but for the Woods-Saxon calculations without (upper panel) and with (lower panel) pairing correlations.

For the case of $^{61}$Zn, the $T=1$ calculations yield minima in the TRS at large deformation, first after the alignment of the neutron $g_{9/2}$ orbits. The excess in the moments of inertia can again be traced back to the sudden drop in the pairing correlations of protons and neutrons. Before the alignment of the neutron $g_{9/2}$, the minimum is very shallow at a smaller deformation, where only a single $g_{9/2}$ orbit is occupied. In contrast, calculations without pairing yield a minimum that is stable over the entire frequency range and has a larger deformation than the one in $^{60}$Zn. Finally, the moments of inertia of $^{59}$Cu are rather flat, however larger than observed in experi-ment. Again, this is due to the decrease in the pairing energy. In $^{59}$Cu, the TRS minimum disappears at low frequencies, and therefore the $T=1$ paired band cannot be followed to low spins.

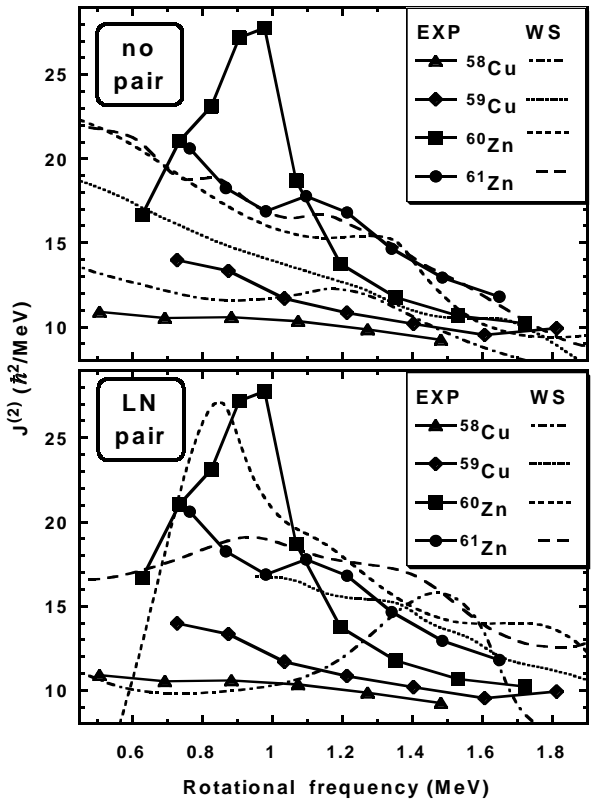

FIG. 6. Same as in Fig. 3 but for the Woods-Saxon calculations without (upper panel) and with (lower panel) pairing correlations.

Since we are dealing with nuclei that are located along the $N=Z$ line, one may pose the question of the role of possible collective $T=0$ pairing and speculate a little about the influence of such correlations. As discussed elsewhere [35], the collective $T=0$ pairing field generally drives the nucleus to somewhat larger deformation, than when only the $T=1$ pair field is present. The sensitivity of our results with respect to the macroscopic model used, may point to either that the $T=1$ field is too strong in our calculations, or by including the $T=0$ field, the results would not be so sensitive to the choice of the macroscopic model. In addition, since the $T=0$ pair field is more resistant at high angular momenta, one may not encounter the unphysical increase in $\mathcal{J}^{(2)}$ that is present in the calculations based on $T=1$ pairing only.

At low spins, the $T=0$ field has essentially the same properties as the $T=1$, i.e., resisting the alignment of quasi-particles. Assuming that part of the correlations in our calculations are indeed due to $T=0$, would not affect much the case of $^{60}$Zn, where we would see a crossing like in the calculations with $T=1$ (possibly shifted to somewhat larger frequencies). However, for the cases of $^{61}$Zn and $^{59}$Cu, the blocking effect would be stronger (due to the n-p blocking), and indeed not much of the alignment would be observed (as is the case in experiment).

To really sort out these intriguing problems, unrestricted calculations need to be performed, that simultaneously take into account both $T=0$ and $T=1$ correlations. We may however already now conclude, that i) in the presence of pairing correlations, one indeed expects a hump in the moment of inertia as is observed for the case of $^{60}$Zn and ii) the simple blocking picture does not hold here, where strong polarizing effects are present, yielding different deformation for the nuclei discussed here and as a result, different pattern of the alignment. Before such complete solutions become available, and the expectations expressed above can be corroborated, in the next section we investigate a very simple non-collective $T=0$ n-p pairing scenario by considering the configuration mixing of unpaired HF solutions.

## IV. HARTREE-FOCK CALCULATIONS WITH THE $T=0$ N-P PAIRING CONFIGURATION MIXING

Apart from the $4^2 4^2$ configuration discussed above, in $^{60}$Zn we also calculated 6 other configurations, namely those that correspond to exciting the $4^2$ proton *and* neutron simultaneously to the negative-parity orbitals $f_\pm$ and $p_\pm$. In principle, there are 16 such excitations possible, however, the lowest ones are obtained by putting the neutron and the proton in *the same* orbitals. This gives 4 configurations denoted by $4^1 f_+ 4^1 f_+$, $4^1 f_- 4^1 f_-$, $4^1 p_+ 4^1 p_+$, and $4^1 p_- 4^1 p_-$. In addition, we also study 2 other configurations obtained by putting the neutron and the proton into the [303]7/2 orbital with *different* signatures, i.e., those denoted by $4^1 f_+ 4^1 f_-$ and $4^1 f_- 4^1 f_+$.

In Fig. 7(b) the energies of the seven configurations selected above are shown with respect to a common rigid-rotor reference energy of $0.025 \times I(I+1)$ MeV. Similarly, Figs. 7(a) and (c) show the analogous configurations in $^{58}$Cu and $^{62}$Ga. Because orbitals $f_\pm$ and $p_\pm$ are very close in energy (cf. Fig. 1), they strongly interact and mix, which very often precludes the convergence of the HF procedure, see discussion in Ref. [36]. Apart from that, the bands of Fig. 7 are shown up to the so-called termination points, i.e., up to the point where the angular-momentum contents of the involved orbitals does not allow for a further angular momentum build up, see Ref. [37], without a significant rearrangement of the nucleons.

By considering the available projections of the total angular momentum $I_y$ for oblate shapes with the $y$ axis as the symmetry axis, one can easily determine the values of the termination-point angular momenta $I_t$, see Table II. The bands obtained in the HF calculations do not always terminate at the oblate axis and can usually be continued beyond $I_t$. However, at angular momenta $I_t$ there always occur significant changes in the structure of bands. Below we discuss and present results only up to the termination points $I_t$.

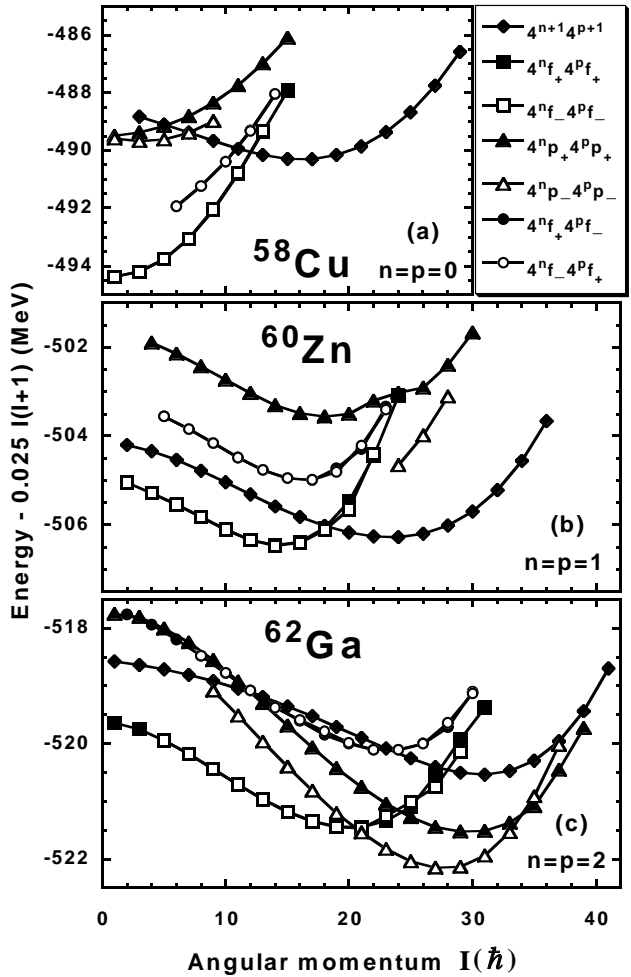

FIG. 7. Energies of selected configurations in $^{58}$Cu (a), $^{60}$Zn (b), and $^{62}$Ga (c), calculated within the HF method with the Skyrme SLy4 interaction, and plotted with respect to a rigid-rotor reference energy. The configurations shown in the legend correspond to the numbers of occupied $N_0$=4 intruder orbitals, $n$ and $p$, that are indicated in each panel.

A conspicuous feature of the HF energies presented in Fig. 7 is the significant energy separation between the n-p paired configurations $4^n f_+ 4^p f_+$ and $4^n f_- 4^p f_-$ on one side, and the broken-pair configurations $4^n f_+ 4^p f_-$ and $4^n f_- 4^p f_+$ on the other side. The former and latter configurations have opposite total signatures, i.e., in the even-even nucleus $^{60}$Zn, configurations $4^1 f_+ 4^1 f_+$ and $4^1 f_- 4^1 f_-$ ($4^1 f_+ 4^1 f_-$ and $4^1 f_- 4^1 f_+$) correspond to $r=+1$ ($r=-1$), while in the odd-odd nuclei $^{58}$Cu and $^{62}$Ga the analogous configurations correspond to $r=-1$ ($r=+1$). Such a signature-separation effect has been for the first time discussed for the SD bands in $^{32}$S [38]. Here it is obtained in the heavier SD region of the $A\simeq 60$ nuclei, as a *mutatis mutandis* identical effect occurring for all the orbitals promoted to the next HO shell.

TABLE II. Values of the termination-point angular momenta $I_t$ (in $\hbar$) for the seven selected configurations in $^{58}$Cu, $^{60}$Zn, and $^{62}$Ga. For convenience, the second column gives the configurations shown in the convention of Refs. [37,17], that, however, does not allow for distinguishing between the signatures of the occupied orbitals.

| Configuration | | $^{58}$Cu $n=p=0$ | $^{60}$Zn $n=p=1$ | $^{62}$Ga $n=p=2$ |
|---|---|---|---|---|
| $4^{n+1} 4^{p+1}$ | $[2(p+1), 2(n+1)]$ | 29 | 36 | 41 |
| $4^n f_+ 4^p f_+$ | $[1p, 1n]$ | 15 | 24 | 31 |
| $4^n f_- 4^p f_-$ | $[1p, 1n]$ | 13 | 22 | 29 |
| $4^n p_+ 4^p p_+$ | $[2p, 2n]$ | 23 | 32 | 39 |
| $4^n p_- 4^p p_-$ | $[2p, 2n]$ | 21 | 30 | 37 |
| $4^n f_+ 4^p f_-$ | $[1p, 1n]$ | 14 | 23 | 30 |
| $4^n f_- 4^p f_+$ | $[1p, 1n]$ | 14 | 23 | 30 |

In Ref. [38] the signature-separation effect was interpreted as a result of the strong n-p attraction transmitted through the time-odd mean fields. Such an attraction is typical for any realistic effective interaction, and it has its origin in the spin-spin components of the interaction. (The signature separation vanishes when in the Skyrme energy functional [39] the coupling constants corresponding to terms $s \cdot s$ and $s \cdot \Delta s$ are set equal to zero.) When averaged within the mean-field approximation, the spin-spin components lead naturally to the time-odd mean fields [39]. Within the phenomenological mean fields, like those given by the Woods-Saxon or Nilsson potentials [40], the time-odd mean fields vanish, and therefore all the four configurations $4^n f_\pm 4^p f_\pm$ are nearly degenerate, i.e., the signature-separation effect occurs only for self-consistent mean fields generated from the spin-spin interactions.

One should note that the four configurations $4^n f_\pm 4^p f_\pm$ have purely independent-particle character (Slater-determinant wave functions), i.e., no collective pair correlations are built into the wave functions. Nevertheless, configurations $4^n f_+ 4^p f_+$ and $4^n f_- 4^p f_-$ contain one more $T$=0 n-p pair as compared to the $4^n f_+ 4^p f_-$ and $4^n f_- 4^p f_+$ configurations, and therefore are sensitive to the n-p pairing component of the effective interaction that is attractive. As a result, the paired configurations $4^n f_+ 4^p f_+$ and $4^n f_- 4^p f_-$ cross the magic configurations $4^{n+1} 4^{p+1}$ at $I$=11, 18, and 27$\hbar$ in $^{58}$Cu, $^{60}$Zn, and $^{62}$Ga, respectively.

The n-p pairing correlations should be, in principle, studied by using methods beyond the mean-field approximation, i.e., by taking into account the configuration-mixing effects for configurations that differ by the n-p pair occupations. The generator-coordinate method (GCM) [40] is the approach of choice for including such effects. It allows for a consistent improvement of wave functions, while staying in the framework of the vari-ational approach. Therefore, the same interaction can be/should be used in the HF method and in the mixing of the HF configurations via the GCM method.

At present, the GCM approach in the rotating frame has not yet been implemented, and in the present study we discuss the same physics problem by introducing a model $T{=}0$ n-p pair-interaction Hamiltonian in the form of

$$\hat{H}_{\text{n-p}} = \hat{H}_0 + \hat{V}_{\text{n-p}} = \sum_{\tau\alpha r}\epsilon_{\tau\alpha r}\hat{N}_{\tau\alpha r} - \sum_{\alpha\beta r} G_{\text{n-p}}^{\alpha\beta}\hat{P}_{\alpha r}^{\dagger}\hat{P}_{\beta r}, \quad (1)$$

where the particle-number ($\hat{N}_{\tau\alpha r}$) and $T{=}0$ n-p pair-creation ($\hat{P}_{\alpha r}^{\dagger}$) operators read

$$\hat{N}_{\tau\alpha r} = a_{\tau\alpha r}^{\dagger} a_{\tau\alpha r}, \quad (2a)$$
$$\hat{P}_{\alpha r}^{\dagger} = a_{\nu\alpha r}^{\dagger} a_{\pi\alpha r}^{\dagger}, \quad (2b)$$

$\tau$ denotes neutrons ($\nu$) or protons ($\pi$), and $\alpha$ and $\beta$ denote the Nilsson labels without the signature quantum number $r$ that is shown explicitly.

Hamiltonian (1) is meant to replace the usual effective-interaction (Skyrme) Hamiltonian when studying the n-p correlation aspects of the nuclear wave functions, and not to be added on top of it. Therefore, the effective single-particle energies $\epsilon_{\tau\alpha r}$ and the coupling constants $G_{\text{n-p}}^{\alpha\beta}=G_{\text{n-p}}^{\beta\alpha}$ have to be angular-momentum and configuration dependent, and Hamiltonian (1) should be understood as a phenomenological interaction operator between configurations that differ by the n-p pair occupations.

The diagonal pairing term can be transformed as

$$\hat{P}_{\alpha r}^{\dagger}\hat{P}_{\alpha r} \equiv \hat{N}_{\nu\alpha r}\hat{N}_{\pi\alpha r}, \quad (3)$$

i.e., it gives a non-zero contribution only if both a neutron and a proton occupy the given $\{\alpha r\}$ orbital. Therefore, the diagonal matrix elements of the n-p pairing Hamiltonian (1) in any given configuration,

$$E(\text{conf.}) = \langle\text{conf.}|\hat{H}_{\text{n-p}}|\text{conf.}\rangle, \quad (4)$$

can be immediately calculated for each Slater determinant. In particular, since in all the four configurations $4^n f_{\pm} 4^p f_{\pm}$ the effective single-particle energies are identical (cf. the Routhian diagram in Fig. 1), the differences of the total energies in the signature-separated configurations read

$$E(4^n f_+ 4^p f_-) - E(4^n f_- 4^p f_+) = 0, \quad (5a)$$
$$E(4^n f_+ 4^p f_+) - E(4^n f_- 4^p f_-) = 0, \quad (5b)$$
$$E(4^n f_+ 4^p f_-) - E(4^n f_+ 4^p f_+) = G_{\text{n-p}}[303]7/2, \quad (5c)$$

where $G_{\text{n-p}}[303]7/2$ stands for the diagonal matrix element $G_{\text{n-p}}^{\alpha\alpha}$ for $\alpha{=}[303]7/2$.

By subtracting the total HF energies of configurations in Eq. (5c), see Fig. 7, one thus obtains an estimate of the n-p pairing diagonal matrix element $G_{\text{n-p}}^{\alpha\alpha}$. Such relative energies (5c) in $^{58}$Cu, $^{60}$Zn, and $^{62}$Ga are plotted in Fig. 8. One can see that the effective matrix elements depend strongly on the angular momentum, and decrease from $G_{\text{n-p}}^{\alpha\alpha}(I{=}0){\simeq}1.6$ ($^{58}$Cu) or 1.9 MeV ($^{60}$Zn and $^{62}$Ga), reaching zero at the termination angular momentum $I_t$. This dependence can be very well parameterized by a simple cubic expression,

$$G_{\text{n-p}}^{\alpha\alpha}(I) = G_{\text{n-p}}^{\alpha\alpha}(I{=}0) \times \left(1 - \left(\frac{I}{I_t}\right)^3\right), \quad (6)$$

shown by dashed lines in Fig. 8.

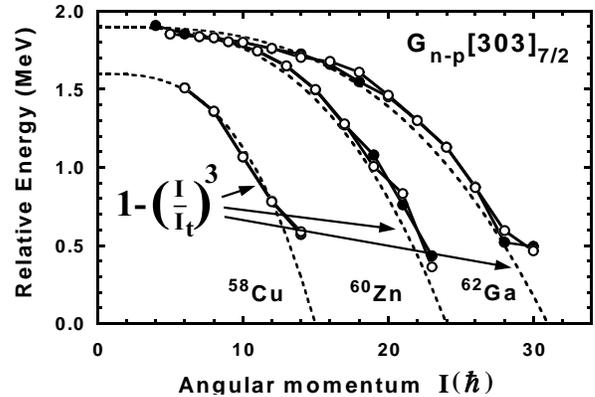

FIG. 8. Energies of configurations $4^n f_+ 4^p f_-$ (closed circles) and $4^n f_- 4^p f_+$ (open circles) in $^{58}$Cu, $^{60}$Zn, and $^{62}$Ga, relative to the corresponding $4^n f_+ 4^p f_+$ configurations. Dashed lines show the simple cubic approximations of Eq. (6). The relative energies can be identified with the angular-momentum-dependent $T{=}0$ n-p pairing matrix elements $G_{\text{n-p}}$ in the [303]7/2 orbital (see text).

A large standard signature splitting of the other single-particle orbitals, which have lower values of the $K$ quantum numbers, does not allow us to determine the other diagonal matrix elements $G_{\text{n-p}}^{\alpha\alpha}$ directly from the HF results, as in Eq. (5c). Of course, such a determination of the non-diagonal matrix elements is not possible either. However, we may use the $I$-dependence of Eq. (6) to postulate a simple separable approximation for the n-p pairing interaction matrix $G_{\text{n-p}}^{\alpha\beta}$ in the form

$$G_{\text{n-p}}^{\alpha\beta}(I) = \sqrt{G_{\text{n-p}}^{\alpha\alpha}(I)G_{\text{n-p}}^{\beta\beta}(I)}. \quad (7)$$

Such a postulate is motivated by the fact that the pairing matrix elements of short-range interactions are given primarily by overlaps between the space wave functions, or more precisely, by the integrals of products of squares of the wave functions. Then, Eq. (7) stems from approximating the integral of products by the product of integrals.

Within the separable approximation (7), the $T{=}0$ n-p pairing interaction $\hat{V}_{\text{n-p}}$ in Hamiltonian (1) takes the simple form of

$$\hat{V}_{\text{n-p}} = -G(I)\left(\hat{P}^\dagger_{+i}\hat{P}_{+i} + \hat{P}^\dagger_{-i}\hat{P}_{-i}\right), \qquad (8)$$

where the $I$-dependent collective n-p pair operators read,

$$\hat{P}^\dagger_r = \sum_\alpha x_\alpha(I)\hat{P}^\dagger_{\alpha r}, \qquad (9)$$

for

$$x_\alpha^2(I) = G_{\text{n-p}}^{\alpha\alpha}(I)/G(I), \qquad (10a)$$

$$G(I) = \sum_\alpha G_{\text{n-p}}^{\alpha\alpha}(I). \qquad (10b)$$

Even then, however, the problem is defined by one parameter per orbital, $G_{\text{n-p}}^{\alpha\alpha}(I=0)$, i.e., it cannot be defined without explicit microscopic configuration-mixing calculations.

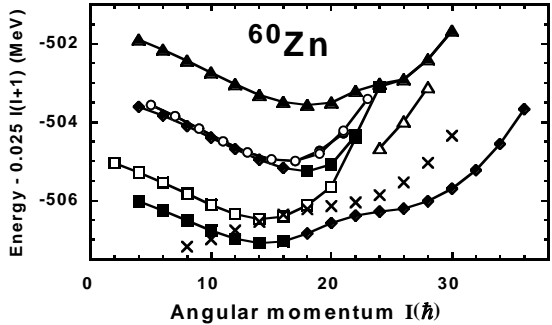

FIG. 9. Same as in Fig. 7(b) but for configurations $4^2 4^2$ and $4^1 f_+ 4^1 f_+$ interacting through the $T=0$ n-p pairing interaction (8) with $G_{\text{n-p}}[303]7/2(I=0)=1.9$ MeV and $G_{\text{n-p}}[440]1/2(I=0)=0.65$ MeV. Crosses show the experimental data in the absolute energy scale. The symbols used in the figure indicate which are the pure configurations of Fig. 7(b) that dominate in the given mixed configurations.

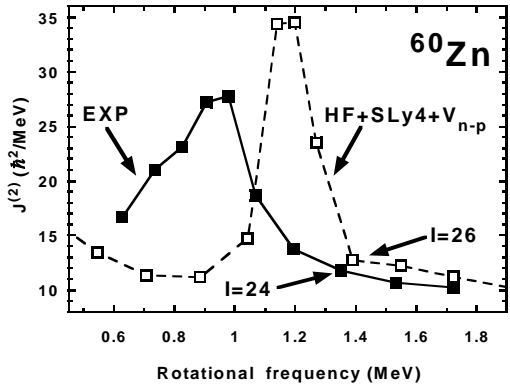

FIG. 10. Experimental (full squares) and calculated (open squares) dynamic moments of inertia $\mathcal{J}^{(2)}$ in the SD band of $^{60}$Zn. Calculations correspond to the lowest band shown in Fig. 9.

Before these become available, in the present study we perform the simplest two-level mixing calculation, in which the two configurations that cross in $^{60}$Zn, $4^2 4^2$ and $4^1 f_+ 4^1 f_+$, see Fig. 7(b), are allowed to interact through the $T=0$ n-p pairing interaction (8). With the diagonal matrix elements of Hamiltonian (1) taken from the HF calculations, and the interaction matrix element defined by the value of $G_{\text{n-p}}[303]7/2(I=0)=1.9$ MeV, also taken from the HF calculations, we are left with one free parameter, i.e., with the value of $G_{\text{n-p}}[440]1/2(I=0)$.

By fixing this parameter at $G_{\text{n-p}}[440]1/2(I=0) = 0.65$ MeV, we obtain at the crossing point of $I=18\hbar$ the effective interaction strength of 0.79 MeV. With the $I$-dependent matrix elements given by Eqs. (6) and (7), we obtain the energies and dynamic moments of inertia shown in Figs. 9 and 10, respectively. It is clear that the mixing and interaction of the $4^2 4^2$ and $4^1 f_+ 4^1 f_+$ configurations correctly reproduces the magnitude of the bump in the $\mathcal{J}^{(2)}$ of $^{60}$Zn.

The position of the crossing point is obtained at frequency or spin that are too large by 0.2 MeV or $4\hbar$, respectively, as compared to experiment. As seen in Fig. 7(b), this position is dictated by the diagonal matrix element $G_{\text{n-p}}[303]7/2$ that shifts down configuration $4^1 f_+ 4^1 f_+$ with respect to the broken-pair degenerate configurations $4^1 f_+ 4^1 f_-$ and $4^1 f_- 4^1 f_+$. As discussed above, such a shift is a direct consequence of the time-odd mean fields resulting from the Skyrme energy density. In the present work we have used the time-odd terms as directly given by the SLy4 Skyrme functional, see Ref. [39], i.e., those that result from fitting the time-even, and not time-odd properties of nuclei. It is clear that a modification of these time-odd terms, that is permitted in the local density approximation, may move the crossing frequency from its current position in Fig. 10. In fact, it is obvious that by decreasing this intensity one may easily decrease the crossing frequency. We do not attempt such a fit here, because the problem of finding good physical values of the time-odd coupling constants is much more general, and it would not make too much sense to make such an adjustment based solely on the specific effect discussed in the present study. We only note in passing that an analogous readjustment of the isovector time-odd coupling constants [41] has led to values that are quite different from those resulting directly from the Skyrme functional.

## V. CONCLUSIONS

In the present study we presented a model description of the four SD bands in nuclei around $^{60}$Zn in order to evaluate the role of the $T=1$ and $T=0$ pairing correlations at high spin in $N \simeq Z$ nuclei. On the one hand, we have shown that calculations with no pairing, whether within the Strutinsky-Woods-Saxon or Skyrme-Hartree-Fock approaches, provide an excellent description of all bands except the one in the doubly-magic $N=Z$ nucleus $^{60}$Zn. On the other hand, the standard Lipkin-Nogami treatment of the $T=1$ pairing gives a fair description of the bump in the second moment of inertia $\mathcal{J}^{(2)}$ in $^{60}$Zn, which results from the simultaneous alignment of the $g_{9/2}$ pairs of neutrons and protons, but fails in describing low values of $\mathcal{J}^{(2)}$ in all the four nuclei at high frequencies. The latter effect results from a gradual disappearance of the $T=1$ pairing correlations with increasing spin, and cannot be avoided if these same pairing correlations have to be responsible for the positive result in $^{60}$Zn. We have also shown that the deformation changes caused by polarization effects of high-$j$ orbitals are strong, and strongly modify the simple blocking picture when going from even to odd isotopes. Nevertheless, even with these polarization effects taken into account in a self-consistent way, the overall description of the discussed set of bands is not satisfactory.

In looking for an alternative physical scenario we have shown that another kind of crossing results from the signature-separation effect that shifts down the HF configurations with neutron-proton pairs present, with respect to those were such pairs are broken. Then the n-p paired configurations cross and interact, giving a correct qualitative reproduction of $\mathcal{J}^{(2)}$ in $^{60}$Zn. This scenario has the advantage that the effect of interaction entirely disappears whenever an odd neutron or an odd proton blocks the n-p pairing interaction. Therefore, non-interacting configurations are obtained in odd neighbors of $^{60}$Zn, yielding a perfect agreement with the data. This result was, however, obtained within a very simple two-band interaction and two-band mixing scheme, while the collective features of the $T=0$ n-p pairing were not studied. Whether or not such collective aspects of a simultaneous presence of the $T=1$ and $T=0$ pairing correlations are important will be the subject of future investigations.

## ACKNOWLEDGMENTS

This research was supported in part by the Polish Committee for Scientific Research (KBN) under Contract No. 5 P03B 014 21, by the French-Polish integrated actions program POLONIUM, and by the computational grants from the *Regionales Hochschulrechenzentrum Kaiserslautern* (RHRK) Germany and from the Interdisciplinary Centre for Mathematical and Computational Modeling (ICM) of the Warsaw University.